\documentclass[12pt]{iopart}
\usepackage{iopams,verbatim}

\def\Journal#1#2#3#4{{#4} {\it #1} {\bf #2} #3 }
\def\CMP{Comm. Math. Phys.}
\def\GRG{Gen. Rel. Grav.}
\def\PRD{Phys. Rev. D}

\begin{document}

\title[Shearfree perfect fluids]{Shearfree perfect fluids with solenoidal magnetic curvature and a $\gamma$-law equation of state}

\author{N Van den Bergh\dag, J Carminati\ddag, H Karimian\dag}
\address{\dag\ Faculty of Applied Sciences TW16, Gent University, Galglaan 2, 9000 Gent, Belgium}
\address{\ddag\ School of Engineering and Information Technology, Deakin University, Australia}
\eads{\mailto{norbert.vandenbergh@ugent.be},\mailto{jcarm@deakin.edu.au},\\
\mailto{Hamidreza.Karimian@UGent.be}}

\begin{abstract}
We show that shearfree perfect fluids obeying an equation of state
$p=(\gamma -1)\mu$ are non-rotating or non-expanding under the
assumption that the spatial divergence of the magnetic part of the
Weyl tensor is zero.

\end{abstract}

%Uncomment for PACS numbers title message

\pacs{04.20.Jb, 04.40.Nr}

\section{Introduction}\label{intro}

The shear-free fluid conjecture claims that for any general
relativistic perfect fluid, in which the energy density $\mu$ and
the pressure $p$ satisfy a barotropic equation of state $p=p(\mu)$
with $p+\mu \not =0$\footnote{otherwise the fluid velocity is not
uniquely determined by the geometry and shearfree, rotating and
expanding examples are known~\cite{Obu}}, necessarily the
expansion $\theta$ or the vorticity $\omega$ vanishes. The
conjecture has been demonstrated in a number of particular cases:
constant $p$ (`dust' with a cosmological
constant)~\cite{Godel,Ellis1,Schucking}, spatial
homogeneity~\cite{Banerji,King}, $d p /d
\mu=1/3$~\cite{Treciokas}, $d p / d \mu= -1/3$ or
$1/9$~\cite{Cyga,VdB}, aligned vorticity and
acceleration~\cite{White}, vanishing magnetic part of the Weyl
tensor ${\bf H}=0$~\cite{Collins}, $\theta =\theta
(\mu)$~\cite{Lang}, $\theta =\theta (\omega )$~\cite{Sopuerta},
Petrov types N~\cite{Carminati1} and III~\cite{Carminati2}, models
in which there is a conformal Killing vector parallel to the fluid
flow~\cite{Coley}, purely magnetic perfect fluids~\cite{Cyga}. It
should be noted that this conjecture is not true in Newtonian
theory and hence if true in general relativity, then it would
highlight essential differences, like that of a well-defined
universal time, between the two theories~\cite{Senovilla}.
Although in most of the above mentioned results a tetrad
formalism was used, recently it has been shown that a covariant
approach can also be successfully employed with considerable
potential for future analysis~\cite{Senovilla,Sopuerta, Chrobok}.

Motivated by the continued absence of a general proof of the
conjecture (or lack of counter-example), we have decided, as an
important step forward, to investigate a direct generalisation of
Collins' 1984 result that $\omega \theta =0$ holds when the
magnetic part $\bf H$ of the Weyl curvature (with respect to the
fluid congruence) vanishes. In our analysis, we shall make the
weaker assumption that $\bf H$ is solenoidal, that is, that its
spatial divergence (see below) vanishes. Interestingly, it has
been recently shown~\cite{Sopuertaetal,VdBetal} that the
assumption of third order restrictions, such as $\textrm{div}
{\bf H} =0$ and/or $\textrm{div} {\bf E}=0$, leads to physically
interesting families of perfect fluid solutions. Also, in a
classification attempt of these fluids the shearfree sub-family
would appear to be a natural first candidate for further
investigation. Although a large part of our analysis holds for a
general barotropic equation of state $p=p(\mu)$ (such as the
positive definiteness of the matrix and the ensuing existence of
a Killing vector parallel to the vorticity), at a certain point,
we shall need to simplify matters by assuming a $\gamma$-law
equation of state (with a possible non-vanishing cosmological term
"effectively" present by allowing $p=(\gamma -1) \mu +
constant$). In the general barotropic case with $D^{b} H_{ab}
=0$, technical difficulties of a very different nature are
present and a number of extra subcases remain to be investigated
(work in progress).

\section{Tetrad choice and relevant equations}\label{outline}
We consider shear-free perfect fluid solutions of the Einstein
field equations
\begin{equation}
R_{ab}-\frac{1}{2}R g_{ab}= \mu u_a u_b +p h_{ab},
\end{equation}
where $\bi u$ is the future-pointing (timelike) unit tangent
vector to the flow, $\mu$ and $p$ are the energy density and
pressure of the fluid and $h_{ab}=g_{ab}+ u_a u_b$ is the
projection tensor into the rest space of the observers with
4-velocity $\bi u$. The vanishing of the shear can be expressed by
\begin{equation}
u_{a;b}= \frac{1}{3}\theta h_{ab}+\omega_{ab}-\dot{u}_a u_b
\end{equation}
where $\theta$ is the (rate of) volume expansion, $\dot{\bi u}$
the acceleration and $\bomega$ is the vorticity.

Throughout, we will assume familiarity with the notations and
conventions of the orthonormal tetrad formalism of
\cite{MacCallum}. Crucial in the successful investigation of any
problem with tetrad formalisms, is the choice of the tetrad
alignment as it can dramatically alter the appearance and
complexity of the resulting complete set of equations. We have
chosen our tetrad as follows. We begin by aligning ${\bi e}_0$
and ${\bi e}_3$  with $\bi u$ and $\bomega$, respectively, such
that $\bomega=\omega {\bi e}_3 \neq 0$. The relevant variables
then become $\mu$, $p$, $\omega$, $\theta$, $\dot{u}_{\alpha}$
and the quantities $\Omega_{\alpha}$, (which determine the
rotation of the triad ${\bi e}_{\alpha}$ with respect to a set of
Fermi propagated axes), together with the quantities $n_{\alpha
\beta}$ and $a_\alpha$. Latin indices will be spacetime indices
while Greek indices will take the values 1, 2 and 3. Further, it
is advantageous to replace $n_{\alpha \beta}$ and $a_\alpha$ with
the new variables $q_{\alpha}, r_{\alpha}$ and $n_{\alpha}$,
defined by
\begin{equation}
n_{\alpha +1\ \alpha-1}=(r_\alpha+q_\alpha)/2,\quad
a_\alpha=(r_\alpha-q_\alpha)/2,\quad n_{\alpha
\alpha}=n_{\alpha+1}+n_{\alpha-1},
\end{equation}
expressions which have to be read modulo 3 (for example $\alpha =
3$ gives $n_{12}= (q_3 + r_3)/2$.) The relation between these
quantities and the Ricci rotation coefficients can be deduced
from the commutators \eref{comms1}. We will also use, as extra
(extension) variables, the components of the spatial gradient of
the expansion
\begin{equation} z_{\alpha} = \partial_{\alpha} \theta \end{equation}
and the (3+1) covariant divergence of the acceleration,
\begin{equation}\label{defj}
j\equiv {\dot{u}^a}_{;a} =
\partial_{\alpha}\dot{u}^{\alpha}+\dot{u}^{\alpha}\dot{u}_{\alpha}-\dot{u}^{\alpha}
(r_{\alpha}-q_{\alpha}).
\end{equation}

As neither the Jacobi identities nor the field equations contain
expressions for the evolution of the $\Omega_{\alpha}$, it is good
practice to choose the triad ${\bi e}_{\alpha}$ such that
$\bOmega+\bomega=0$. The fact that this is always possible follows
from the results of the action of the commutators $[\partial_3,\,
\partial_1]$ and $[\partial_3,\, \partial_2]$ on $p$. One finds
that $\Omega_1=\Omega_2=0$ after which a rotation in the
(12)-plane can be chosen such that $\Omega_3 +\omega=0$. Herewith,
the tetrad is determined up to rotations in the (12)-plane over
an angle $\alpha$ satisfying $\partial_0 \alpha =0$. Noticing
that the evolution equations for the quantities $n_{11}-n_{22}$
and $n_{12}$ are identical (see \cite{MacCallum}), allows one to
further fix the tetrad by making either $n_{11}-n_{22}=0$ or
$n_{12} = 0$. Henceforth our choice will be
\begin{equation}n_{11}=n_{22}\equiv n.
\end{equation}

\noindent We express the vanishing of the spatial divergence of
the magnetic part of the Weyl tensor, $H_{ab} = C_{acbd} u^c u^d
$, by means of the Bianchi identity~\cite{MaBa},
\begin{equation}
(\textrm{div} H)_a\equiv D^b H_{ab} =3 \omega^{b} E_{ab} +(\mu+p)
\omega_{a}.
\end{equation}
Here the spatial derivative operator is defined by
$$D_a{S^{c\ldots d}}_{e\ldots f} = {h_a}^b{h^c}_p\cdots
{h^{d}}_{q}{h_{e}}^{r}\cdots {h_{f}}^{s} \nabla_b{S^{p\ldots
q}}_{r\ldots s}.$$

\noindent This shows that when $D^{b} H_{ab} =0$, the vorticity is
an eigenvector of the electric part of the Weyl tensor, $E_{ab}=
C_{acbd} u^c u^d$, with eigenvalue $-(\mu+p)/3$. With our choice
of tetrad this implies
\begin{eqnarray}\label{divH}
E_{13}=E_{23}=0, \label{divH} \\
E_{33}= -\frac{\mu+p}{3}\label{divHb}.
\end{eqnarray}
In order to relate the components of $E_{ab}$ with the spatial
gradient of the acceleration, we will make use of the Ricci
identity~\cite{MaBa}
\begin{equation}\label{defE}
E_{ab} = D_{\langle a }\dot{u}_{b\rangle}- \omega_{\langle
a}\omega_{b\rangle} + \dot{u}_{\langle a}\dot{u}_{b\rangle},
\end{equation}
where $S_{\langle a b \rangle}$ stands for the spatially projected
and trace-free part of $S_{ab}$.

The basic equations of the formalism are now the Einstein field
equations and the Jacobi equations, which we present, using the
simplifications above, in the appendix.

\noindent Henceforth we will assume $\gamma \neq
1$~\cite{Godel,Schucking,Ellis1},
$\gamma\neq\frac{4}{3}$~\cite{Treciokas},
$\gamma\neq\frac{2}{3}$~\cite{VdB},
$\gamma\neq\frac{10}{9}$~\cite{VdB} and, of course $\gamma \neq 0$
(cf.~introduction).

We shall argue by contradiction in order to establish that a
shearfree fluid under the given conditions satisfies $\omega
\theta =0$. The important step is contained in section 3, where we
prove first that, if $\omega \theta \neq 0$, a Killing vector
exists parallel to the vorticity. This results in a great
simplification of the governing equations and, although this
subcase still contains all the intricacies of the `general'
problem, it allows us to complete the proof in section 4.

\section{Proof of the existence of a Killing vector parallel to
$\bomega$ for $\bomega\btheta \neq 0$}
 First notice that the equations
(\ref{j8},\ref{j14}) immediately lead to evolution equations for
the variables $r_{\alpha}$ and $q_{\alpha}$,
\begin{eqnarray}
3 \partial_0 r_{\alpha}   = - \theta \dot{u}_{\alpha} - z_{\alpha}- \theta r_{\alpha}  \label{j8_14a}\\
3 \partial_0 q_{\alpha} = z_{\alpha} + \theta (\dot{u}_{\alpha}
+q_{\alpha}) \label{j8_14b}
\end{eqnarray}
while (\ref{j1}) and the $(0\alpha)$ field equations
(\ref{ein01}-\ref{ein03}) give us the spatial derivatives of
$\omega$
\begin{eqnarray}
\partial_1 \omega = \frac{2}{3} z_2-\omega(q_1+2 \dot{u}_1)\label{gradom1}\\
\partial_2 \omega = - \frac{2}{3} z_1+\omega (r_2-2\dot{u}_2)\label{gradom2}\\
\partial_3 \omega = \omega( \dot{u}_3+r_3-q_3)\label{gradom3}
\end{eqnarray}
together with the algebraic restriction
\begin{equation}\label{defn33}
n_{33} = \frac{2}{3\omega} z_3.
\end{equation}
The evolution equation for $n$ follows from (\ref{j11}):
\begin{equation}
\partial_0 n = -\frac{\theta}{3} n.
\end{equation}
Acting with the commutators $[\partial_0,\,
\partial_{\alpha}]$ and $[\partial_1,\, \partial_2]$  on the pressure and using (\ref{j7}) together with the conservation laws
\begin{eqnarray}
\partial_0 \mu =-\theta (\mu +p)\label{cons} \\
\partial_{\alpha} p = -(\mu +p) \dot{u}_{\alpha}\label{consb}
\end{eqnarray}
leads to  a first set of evolution equations for the acceleration
and vorticity:
\begin{equation} \eqalign{
\partial_0 \dot{u}_{\alpha} = (\gamma -1) z_{\alpha} + (\gamma-\frac{4}{3}) \theta \dot{u}_{\alpha},\\
\partial_0 \omega =(\gamma-\frac{5}{3}) \omega \theta .}
\end{equation}
The spatial derivatives of the acceleration can be obtained from
(\ref{defE}), using (\ref{defj}):
\begin{eqnarray}
\partial_{1}\dot{u}_{1} = -\frac{1}{3} \omega^2+\frac{1}{3} j-\dot{u}_2 q_2+\dot{u}_3 r_3-\dot{u}_1^2+E_{11}\label{e1u1}\\
\partial_{2}\dot{u}_{2} = -\frac{1}{3} \omega^2+\frac{1}{3} j-\dot{u}_3 q_3+\dot{u}_1 r_1-\dot{u}_2^2+E_{22}\label{e2u2}\\
\partial_{3}\dot{u}_{3} = -\frac{1}{3} \omega^2+\frac{1}{3} j-\dot{u}_1 q_1+\dot{u}_2 r_2-\dot{u}_3^2+E_{33}\label{u33}\\
\partial_1 \dot{u}_2 = \omega \theta(1-\gamma)+q_2 \dot{u}_1+\frac{1}{2} n_{33} \dot{u}_3-\dot{u}_1 \dot{u}_2+E_{12}\label{e1u2}\\
\partial_2 \dot{u}_1 = -\omega \theta(1-\gamma)-r_1 \dot{u}_2-\frac{1}{2} n_{33} \dot{u}_3-\dot{u}_1 \dot{u}_2+E_{12}\label{e2u1}\\
\partial_1 \dot{u}_3 = -r_3 \dot{u}_1-\frac{1}{2} \dot{u}_2 n_{33}-\dot{u}_1 \dot{u}_3+E_{13} \label{e1u3} \\
\partial_2 \dot{u}_3 = \frac{1}{2} \dot{u}_1 n_{33}+q_3 \dot{u}_2-\dot{u}_2 \dot{u}_3+E_{23} \label{e2u3} \\
\partial_3 \dot{u}_1 = -\frac{1}{2} \dot{u}_2 n_{33}+n \dot{u}_2+q_1 \dot{u}_3-\dot{u}_1 \dot{u}_3+E_{13}\\
\partial_3 \dot{u}_2 = \frac{1}{2} \dot{u}_1 n_{33}-n \dot{u}_1-r_2 \dot{u}_3-\dot{u}_2 \dot{u}_3+E_{23}
\end{eqnarray}
Next we act with the $[\partial_0,\,
\partial_{\alpha}]$ commutators on $\omega$ and $\theta$ and use the propagation of (\ref{defn33}) along $\bi u$ in order to
obtain expressions for the evolution of $z_{\alpha}$ and the
spatial gradient of $j$:
\begin{equation}\label{Zevol}
\eqalign{
\partial_0 z_1 = \theta (\gamma-2) z_1-\frac{9\gamma-10}{2}\omega \left(z_2+\theta \dot{u}_2\right) \label{e0Z1}\\
\partial_0 z_2 = \theta (\gamma-2) z_2+\frac{9\gamma-10}{2}\omega \left(z_1+\theta \dot{u}_1\right) \label{e0Z2} \\
\partial_0 z_3 = \theta (\gamma-2) z_3 \label{e0Z3}
}
\end{equation}
and
\begin{eqnarray}
\partial_1 j &=& \theta z_1 (\gamma-1)-z_2 \frac{-14+27 \gamma}{6} \omega-\dot{u}_1 j-\frac{-1+2 \gamma}{2(\gamma-1)} \dot{u}_1 \mu-\frac{\gamma-1}{2}
\dot{u}_1 p\nonumber \\
& & +\frac{1}{3} \dot{u}_1 (18 \omega^2+\theta^2)-\dot{u}_2 \frac{9 \gamma-10}{2} \theta \omega+4 \omega^2 q_1 \label{e1j}\\
\partial_2 j &=& \theta z_2 (\gamma-1)+z_1 \frac{-14+27 \gamma}{6} \omega-\dot{u}_2 j-\frac{-1+2 \gamma}{2 (\gamma-1)} \dot{u}_2 \mu-\frac{\gamma-1}{2}
\dot{u}_2 p\nonumber \\
& & +\frac{1}{3} \dot{u}_2 (18 \omega^2+\theta^2)+\dot{u}_1 \frac{9 \gamma-10}{2} \theta \omega-4 \omega^2 r_2 \label{e2j}\\
\partial_3 j &=& \theta z_3 (\gamma-1)-\dot{u}_3 j-\frac{-1+2 \gamma}{2(\gamma-1)} \dot{u}_3 \mu-\frac{\gamma-1}{2} \dot{u}_3 p+\frac{1}{3} \dot{u}_3
(-18 \omega^2+\theta^2) \nonumber \\
& &-4 (r_3-q_3) \omega^2 .
\end{eqnarray}
Now we may evaluate $\sum_{\alpha}[\partial_0,\,
\partial_{\alpha}]\dot{u}_{\alpha}$ by using (\ref{defj}) and
(\ref{e1u1}-\ref{u33}) and this leads to the evolution of $j$ as
\begin{eqnarray}\label{e0jfirst}
\partial_0 j &=& (\gamma-1) \partial_{\alpha}
z^{\alpha}+\theta (\gamma-\frac{5}{3}) j +(2\gamma -\frac{5}{3})
\theta \dot{u}_{\alpha} \dot{u}^{\alpha}+(4\gamma -\frac{11}{3})
\dot{u}_{\alpha}z^{\alpha}\nonumber \\
& & +(\gamma-1)(q_{\alpha}-r_{\alpha})z^{\alpha}
\end{eqnarray}

\noindent Next a long calculation, involving the propagation of
the field equations (\ref{ein12}-\ref{ein33}) along $\bi u$ and
the use of the $[\partial_0,\, \partial_{\alpha}]$ commutators on
$r_{\beta}, q_{\beta}, \dot{u}_{\beta}$ together with the
$[\partial_{\alpha},\,
\partial_{\beta}]$ commutators on $\theta$ and $\omega$, allows one
to obtain algebraic expressions for the directional derivatives
$\partial_{\alpha} z_{\beta}$, as well as evolution equations for
$j$ and $E_{\alpha \beta}$ (the latter can equally well be
obtained by writing out the `dot E' Bianchi
identities~\cite{MaBa}). The resulting expressions can be
simplified by the introduction of the following non-linear
combinations of kinematic quantities, each having a clear
geometric meaning,
\begin{equation}
U= \dot{u}_1^2+\dot{u}_2^2, V=\dot{u}_1 z_1 +\dot{u}_2 z_2,
W=\dot{u}_1 z_2-\dot{u}_2 z_1, Z=z_1^2+z_2^2 .
\end{equation}
We obtain:\\

\noindent a) the evolution of $j$:

\begin{equation}\label{e0j}
\fl \partial_0 j  = \frac{6 \gamma-5}{3}(V+\theta U+z_3
\dot{u}_3+\theta \dot{u}_3^2)+(\gamma-1)(9 \gamma-10)\omega^2
\theta+\frac{3 \gamma-5}{3} j \theta;
\end{equation}
b) the evolution of $E_{\alpha \beta}$:
\begin{eqnarray}
\fl \partial_0 E_{11} = -\frac{2}{3} \theta E_{11}\nonumber
\\ +\frac{2(3\gamma-4)}{9(3\gamma-2)}(2 \dot{u}_1  z_1 -\dot{u}_2 z_2-
\dot{u}_3 z_3 + (\omega^2 (9 \gamma-8)+2 \dot{u}_1^2-\dot{u}_2^2-\dot{u}_3^2) \theta),\\
\fl \partial_0 E_{22} = -\frac{2}{3} \theta E_{22} \nonumber
\\
-\frac{2(3\gamma-4)}{9(3\gamma-2)}(\dot{u}_1 z_1-2 \dot{u}_2 z_2 +
\dot{u}_3 z_3
- (\omega^2 (9 \gamma-8)-\dot{u}_1^2+2 \dot{u}_2^2-\dot{u}_3^2) \theta),\\
\fl \partial_0 E_{12} = -\frac{2}{3} \theta E_{12}+
\frac{3\gamma-4}{3(3\gamma-2)} (\dot{u}_2 z_1+
\dot{u}_1 z_2+ 2 \dot{u}_1 \dot{u}_2 \theta),\\
\fl \partial_0 E_{13} = -\frac{2}{3} \theta E_{13}+
\frac{3\gamma-4}{3(3\gamma-2)}( \dot{u}_3  z_1+ \dot{u}_1
z_3+2\dot{u}_1 \dot{u}_3  \theta) \label{e0E13},\\
\fl \partial_0 E_{23} = -\frac{2}{3}\theta
E_{23}+\frac{3\gamma-4}{3(3\gamma-2)} (\dot{u}_3  z_2+\dot{u}_2
z_3+2\dot{u}_2 \dot{u}_3 \theta)\label{e0E23};
\end{eqnarray}
c) the spatial derivatives of $z_{\alpha}$:
\begin{eqnarray}
\fl \partial_1 z_1  = -\frac{2 (15 \gamma-14)}{3(3 \gamma-2)}
\dot{u}_1 z_1+\frac{\dot{u}_2 (6  \gamma-8)+q_2(6-9 \gamma)}{3(3
\gamma-2)} z_2
+\frac{\dot{u}_3 (6 \gamma-8)+r_3(9 \gamma-6)}{3(3 \gamma-2)} z_3\nonumber \label{e1Z1}\\
 \fl +\frac{\theta}{3(3 \gamma-2)}
  (\omega^2(2-33 \gamma+27 \gamma^2)-(6 \gamma-8) (2 \dot{u}_1^2-\dot{u}_2^2-\dot{u}_3^2 )+(6-9 \gamma) E_{11}) \label{Z11},\\
\fl \partial_2 z_2  = -\frac{2(15 \gamma-14)}{3(3 \gamma-2)}
\dot{u}_2 z_2 +\frac{\dot{u}_1 (6  \gamma-8)+r_1 (9 \gamma-6)}{3(3
\gamma-2)} z_1
+\frac{\dot{u}_3 (6 \gamma-8)-q_3(9 \gamma-6)}{3(3 \gamma-2)} z_3\nonumber \label{e2Z2}\\
 \fl +\frac{\theta}{3(3 \gamma-2)}
(\omega^2(2-33 \gamma+27 \gamma^2)+(6 \gamma-8) (\dot{u}_1^2-2\dot{u}_2^2+\dot{u}_3^2) +(6-9 \gamma) E_{22})\label{Z22},\\
\fl \partial_3 z_3  = -\frac{2(15 \gamma-14)}{3(3 \gamma-2)}
\dot{u}_3 z_3+\frac{\dot{u}_1 (6 \gamma-8)+q_1(6-9 \gamma)}{3(3
\gamma-2)} z_1+\frac{\dot{u}_2 (6 \gamma-8)-r_2(6-9 \gamma)}{3(3
\gamma-2)} z_2
\nonumber \\
 \fl +\frac{\theta}{3(3 \gamma-2)} (
\omega^2(56-78 \gamma+27 \gamma^2)+(6 \gamma-8) U +(16-12 \gamma) \dot{u}_3^2+(6-9 \gamma) E_{33}\label{Z33},\\
\fl \partial_1 z_2 = q_2 z_1 -\frac{6 (\gamma-1
)}{3\gamma-2}(\dot{u}_2 z_1+ \dot{u}_1 z_2)+\frac{1}{2} n_{33}
z_3-\frac{2(3\gamma-4)}{3\gamma-2} \dot{u}_1 \dot{u}_2
\theta-\theta E_{12}\nonumber \\ \fl +\omega(\frac{\mu+3 p}{2}-j
+\frac{1}{3} \theta^2-2 \omega^2)\label{e1Z2},\\
\fl \partial_2 z_1 = -r_1 z_2 -\frac{6 (\gamma-1
)}{3\gamma-2}(\dot{u}_2 z_1+ \dot{u}_1 z_2)-\frac{1}{2} n_{33}
z_3-\frac{2(3\gamma-4)}{3\gamma-2} \dot{u}_1 \dot{u}_2
\theta-\theta E_{12}\nonumber \\ \fl -\omega(\frac{\mu+3 p}{2}-j
+\frac{1}{3} \theta^2-2 \omega^2)\label{e2Z1},\\
\fl \partial_3 z_1 = q_1 z_3-\frac{1}{2} (n_{33}-2 n)z_2 -\frac{6
(\gamma-1 )}{3\gamma-2}(\dot{u}_3 z_1+ \dot{u}_1
z_3)-\frac{2(3\gamma-4)}{3\gamma-2} \dot{u}_1
\dot{u}_3\theta-\theta E_{13},
\\
\fl \partial_3 z_2 = -r_2 z_3+\frac{1}{2} (n_{33}-2 n)z_1 -\frac{6
(\gamma-1 )}{3\gamma-2}(\dot{u}_3 z_2+ \dot{u}_2
z_3)-\frac{2(3\gamma-4)}{3\gamma-2} \dot{u}_2
\dot{u}_3\theta-\theta E_{23},
\\
\fl \partial_1 z_3 = -r_3 z_1-\frac{1}{2} n_{33}z_2 -\frac{6
(\gamma-1 )}{3\gamma-2}(\dot{u}_3 z_1+ \dot{u}_1
z_3)-\frac{2(3\gamma-4)}{3\gamma-2} \dot{u}_1
\dot{u}_3\theta-\theta E_{13}\label{e1Z3},
\\
\fl \partial_2 z_3 = q_3 z_2+\frac{1}{2} n_{33}z_1 -\frac{6
(\gamma-1 )}{3\gamma-2}(\dot{u}_3 z_2+ \dot{u}_2
z_3)-\frac{2(3\gamma-4)}{3\gamma-2} \dot{u}_2
\dot{u}_3\theta-\theta E_{23}\label{e2Z3}.
\end{eqnarray}

\noindent The evolution of the new variables $U$, $V$, $W$ and $Z$
is given by
\begin{equation}\eqalign{
\partial_0 U = \frac{2(3 \gamma-4)}{3} \theta U +2 (\gamma-1) V\\
\partial_0 V = \frac{2(3 \gamma -5)}{3}\theta  V-\frac{9 \gamma-10}{2} \omega  W +(\gamma-1) Z\\
\partial_0 W =\frac{9 \gamma-10}{2}\theta\omega  U+\frac{9 \gamma-10}{2} \omega V+\frac{2(3
\gamma-5)}{3}\theta W\\
\partial_0 Z = \theta \omega (9 \gamma-10) W+2 (\gamma-2) \theta
Z}.
\end{equation}

The equations given so far describe the behaviour of a general
shearfree perfect fluid with a $\gamma$-law equation of state (and
hence can be useful in further investigations of the $\omega
\theta =0$ conjecture). \underline{It is only at this stage that
we introduce the condition that the} \\
\underline{magnetic part of the Weyl tensor is divergence-free.}
Imposing (\ref{divH}, \ref{divHb}) we deduce from (\ref{e0E13},
\ref{e0E23})
\begin{eqnarray}
(\gamma-\frac{4}{3}) (z_3+2\theta\dot{u}_3) \dot{u}_1+\dot{u}_3 z_1= 0 \label{main1}\\
(\gamma-\frac{4}{3}) (z_3+2\theta\dot{u}_3) \dot{u}_2+\dot{u}_3
z_2=0, \label{main2}
\end{eqnarray}
where the factor $\gamma-\frac{4}{3}$ can be divided out
($\gamma=\frac{4}{3}$ corresponds to a pure radiation perfect
fluid, in which case the conjecture $\omega \theta =0$ is known to
hold\cite{Treciokas}).

\subsection{$z_3+2\theta\dot{u}_3\neq 0$}

First suppose that $z_3+2\theta\dot{u}_3\neq 0$. Now propagating
equations (\ref{main1},\ref{main2}) along $\bi u$ results in a
linear system in $z_1, z_2$,
\begin{equation}\label{main3}
\alpha z_1+\beta z_2=0, \  -\beta z_1 +\alpha z_2 =0
\end{equation}
with \begin{equation}\eqalign{
\alpha=(\gamma-1)(z_3+\dot{u}_3 \theta)^2+\left(\frac{\mu+3 p}{2}- j+(\gamma -\frac{4}{3})\theta^2-2\omega^2 \right) \dot{u}_3^2\\
\beta=-\frac{9 \gamma -10}{4}(z_3+\theta\dot{u}_3)\omega
\dot{u}_3. }
\end{equation}
If $\beta \neq 0$ the system (\ref{main3}) has only the
0-solution, and consequently (\ref{main1},\ref{main2}) would imply
$\dot{u}_1=\dot{u}_2=0$. In this case, the acceleration would be
parallel to the vorticity and the conjecture $\omega \theta = 0$
follows by \cite{White}. When $\beta =0$ we have the following
possibilities:
\begin{itemize}
\item $\gamma=\frac{10}{9}$: in this case the
conjecture $\omega \theta = 0$ has been demonstrated~\cite{VdB}
\item  $\dot{u}_3=0$: (\ref{main1}) now implies $\dot{u}_1=\dot{u}_2=0$ which takes us back to the `dust' cases with cosmological constant~\cite{Ellis1,Godel,Schucking}
\item $z_3+\theta \dot{u}_3=0$ and hence, by (\ref{main1}), $\dot{u}_3(z_1+\theta \dot{u}_1)=\dot{u}_3(z_2+\theta \dot{u}_2)=0$.
Assuming $\dot{u}_3\neq 0$ this implies that the spatial gradient
of $\log \theta -\int \frac{d p}{\mu+p}$ vanishes, and the fluid
flow is irrotational (unless of course $\log \theta -\int \frac{d
p}{\mu+p}=0$, in which case the $\omega \theta=0$ conjecture
would follow from \cite{Lang}), which contradicts our assumption.
\end{itemize}

\subsection{$z_3+2\theta\dot{u}_3 = 0$}\label{S:kil2}
If $z_3+2\theta\dot{u}_3 = 0$ then (\ref{main1}) implies that
$\dot{u}_3 z_1 =\dot{u}_3 z_2=0$. If $\dot{u}_3\neq 0$ propagation
of the conditions $z_1=z_2=0$ along $\bi u$ gives, using
(\ref{Zevol}),
\begin{equation}
\dot{u}_1 (9 \gamma -10) \omega\theta = \dot{u}_2 (9 \gamma -10)
\omega\theta = 0
\end{equation}
which again would imply that the vorticity and acceleration are
parallel. We therefore conclude that $\dot{u}_3=0$ and hence also
$z_3=0$. By (\ref{ein03}) one also has then $n_{33}=0$.
Furthermore, with (\ref{e1u3},\ref{e2u3}) and
(\ref{e1Z3},\ref{e2Z3}) the conditions $\dot{u}_3=z_3=0$ yield
\begin{equation}\eqalign{
r_3 z_1 = q_3 z_2=0,\  r_3 \dot{u}_1 = q_3 \dot{u}_2 =0,}
\end{equation}
such that, unless acceleration and vorticity are parallel, at
least one of $q_3$ or $r_3=0$. Without loss of generality we can
suppose $q_3=0$. If $r_3\neq0$ then $z_1=\dot{u}_1=0$, after which
(\ref{e0Z1}) would imply $(9 \gamma-10) \omega (z_2+\theta
\dot{u}_2)=0$. The spatial gradients of $\theta$ and $p$ would
then be parallel (with the coefficient of proportionality depending
on $\theta$ only), which again would imply that the flow is
irrotational. We conclude that both $q_3=r_3=0$. A straightforward
calculation shows then that there is a Killing vector along
$\mathbf{e}_3$, although in the sequel this property will not be
explicitly used.

\section{Proof of $\bomega \btheta =0$}
From (\ref{u33},\ref{Z33}) and $\dot{u}_3=z_3=0$ one obtains
\begin{eqnarray}
\fl j+2\omega^2-3\dot{u}_1 q_1+3 \dot{u}_2 r_2+3 E_{33} = 0 \label{vgl1}\\
\fl (3 \gamma -2)(-3 q_1 z_1+3 r_2 z_2-3 \theta E_{33})+ (3 \gamma
-4) (2 \theta\ U+2 V+(9\gamma-14)\omega^2\theta )=0 \label{vgl2}
\end{eqnarray}
Propagating (\ref{divHb}) along $\bi u$ gives a third algebraic
relation among the same variables, which allows us to express
$q_1$ and $r_2$ as
\begin{eqnarray} \fl q_1=
\left((\gamma-1)(p+\mu)+3(3\gamma-4)\omega^2\right)\theta
\dot{u}_2
W^{-1}+(j-p-\mu+2 \omega^2) z_2 (3 W)^{-1} \label{q1_}\\
\fl r_2= \left((\gamma-1)(p+\mu)+3(3\gamma-4)\omega^2\right)\theta
\dot{u}_1 W^{-1}+(j-p-\mu+2 \omega^2) z_1 (3 W)^{-1}\label{r2_}
\end{eqnarray}
and by which we can rewrite (\ref{vgl2}) as
\begin{equation}
(3 \gamma-2)^2 (p+\mu)\theta+(3\gamma-4)
\left(4(9\gamma-8)\theta\omega^2+2 V+2\theta U\right)=0.
\end{equation}
Note that $W\neq0$ since propagating $W=0$ along $\bi u$ would
imply $(9 \gamma-10) U \omega (z_1+\theta\dot{u}_1)=0$ and hence
$z_1+\theta \dot{u}_1=z_2+\theta \dot{u}_2=0$. Applying the same
reasoning as in section \ref{S:kil2} above would then lead to an
irrotational flow.

We now focus on equations (\ref{e1j},\ref{e2j}). Applying the
$[\partial_1,\,
\partial_2]$ commutator on $j$ and using (\ref{gradom1}, \ref{gradom2},
\ref{e1u1}, \ref{e2u2}, \ref{e1u2}, \ref{e2u1},\ref{e1Z1},
\ref{e2Z2}, \ref{e1Z2}, \ref{e2Z1}, \ref{j4}, \ref{ein12}) we
obtain
\begin{equation}\label{vgl7} \fl 2 (4\gamma-5) W+(7\gamma-10)\omega
j+(3\gamma-2)(11\gamma-8)(p+\mu)\omega+(459\gamma^2-1048\gamma+580)\omega^3=0.
\end{equation}
Propagating this relation twice along $\bi u$ and simplifying the
result using (\ref{vgl2}) and (\ref{vgl7}) results in a linear
system
\begin{equation} \fl
\left[
  \begin{array}{cc}
    6 \gamma (21 \gamma-20) (3 \gamma-4) &
(3 \gamma-2) (47 \gamma^2-76 \gamma+30)  \\
    4 (5-3 \gamma)(21 \gamma-20) (3
\gamma-4)  &
(3 \gamma-2) (47 \gamma^2-76 \gamma+30) \\
  \end{array}
\right] \left[
          \begin{array}{c}
            \omega^2 \\
            p+\mu \\
          \end{array}
        \right]
 =0,
\end{equation}
which clearly implies $\omega=0$ or $p+\mu=0$.

\section{Discussion}
We have demonstrated that a shearfree perfect fluid, obeying an
equation of state $p=(\gamma -1) \mu$ and satisfying the
`solenoidal' condition $\textrm{div} H =0$, is either expansion
free or vorticity free. Our proof relies heavily on the subcases
discussed earlier in the literature for various values of
$\gamma$. In addition, we have presented an interesting framework
which shows promise for successfully mounting an assault on the
shearfree conjecture with the assumption of just a gamma law.
Presently we are trying to generalise our result to include an
arbitrary barotropic equation of state $p=p(\mu)$ with the
solenoidal condition. Preliminary work shows that almost all
relations in the present work can be generalised (in particular
the fact that $\omega \theta \neq 0$ would imply the existence of
a Killing vector parallel to the vorticity). Some technical
difficulties still remain to be dealt with in the final stage of
the proof. This is primarily due to the resulting more complicated
nature of equation $(\ref{vgl7})$ and its subsequent derivatives. A
proof of the conjecture would be very desirable, as the
classification of the shearfree case would form a necessary and
natural first step in the study of purely `solenoidal' perfect
fluids.

\section{Appendix}

Commutator relations, using $\sigma_{\alpha \beta}=0$:
\begin{equation}
\eqalign{ \fl [\partial_0 , \partial_1] = {\dot u}_1 \partial_0
-\theta_1 \partial_1 +(\omega_3+\Omega_3) \partial_2
-(\omega_2+\Omega_2) \partial_3 \label{comms1}\\
\fl [\partial_0 , \partial_2] = {\dot u}_2 \partial_0
-(\omega_3+\Omega_3) \partial_1 -\theta_2 \partial_2
+(\omega_1+\Omega_1) \partial_3 \label{comms2}\\
\fl [\partial_0 , \partial_3] = {\dot u}_3 \partial_0
+(\omega_2+\Omega_2) \partial_1 -\theta_3 \partial_3
-(\omega_1+\Omega_1) \partial_2 \label{comms3}\\
\fl [\partial_1 , \partial_2] = -2 \omega_3 \partial_0 + q_2
\partial_1
+r_1 \partial_2 +n_{33} \partial_3 \label{comms4}\\
\fl [\partial_2 , \partial_3] =  -2 \omega_1 \partial_0 + q_3
\partial_2
+r_2 \partial_3 +n_{11} \partial_1 \label{comms5}\\
\fl [\partial_3 , \partial_1] = -2 \omega_2 \partial_0 + r_3
\partial_1 +q_1 \partial_3+n_{22} \partial_2 \label{comms6}}
\end{equation}\ \\

\noindent Using in addition the simplifications
$\omega_1=\omega_2=\Omega_1=\Omega_2=0$, $\Omega_3 = - \omega_3$,
$n_{11}=n_{22}=n$ one obtains the:\\

\noindent Jacobi equations:
\begin{eqnarray}
\fl \partial_3 \omega  = \omega (\dot{u}_3+r_3-q_3) \label{j1} \\
\fl \partial_1 n + \partial_2 r_3  + \partial_3 q_2  - n (r_1 - q_1 ) - r_3 r_2 + q_3 q_2 = 0 \label{j2}\\
\fl \partial_2 n  + \partial_3 r_1  + \partial_1 q_3  - n (r_2 - n q_2) - r_3 r_1 + q_3 q_1  = 0 \label{j3}\\
\fl \partial_1 r_2 +\partial_2 q_1 +\partial_3 n_{33} -\frac{2}{3} \theta \omega-r_2 r_1+q_2 q_1-n_{33}( r_3- q_3) = 0 \label{j4}\\
\fl \partial_2 \dot{u}_3  - \partial_3 \dot{u}_2  - n \dot{u}_1 - q_3 \dot{u}_2  -  r_2 \dot{u}_3 = 0 \label{j5}\\
\fl \partial_3 \dot{u}_1  - \partial_1 \dot{u}_3  - n \dot{u}_2 - r_3 \dot{u}_1   - q_1 \dot{u}_3  = 0 \label{j6}\\
\fl 2 \partial_0 \omega +\partial_1 \dot{u}_2 -\partial_2 \dot{u}_1 -\dot{u}_1 q_2-\dot{u}_2 r_1- \dot{u}_3 n_{33}+\frac{4}{3} \theta \omega = 0 \label{j7}\\
\fl \partial_0 (r_{\alpha}-q_{\alpha})-\frac{1}{3}\partial_{\alpha} \theta +z_{\alpha}+\frac{\theta}{3}(r_{\alpha}-q_{\alpha}+2 \dot{u}_{\alpha}) = 0 \label{j8}\\
\fl 3 \partial_0 n  + 3 \partial_3 \omega  + n \theta - 3 \omega (\dot{u}_3 + r_3- q_3) = 0 \label{j11}\\
\fl 3 \partial_0 n_{33} + 3 \partial_3 \omega + n_{33} \theta - 3 \omega (\dot{u}_3  + r_3 - q_3) = 0 \label{j13}\\
\fl \partial_0
(r_{\alpha}+q_{\alpha})+\frac{1}{3}\theta(r_{\alpha}+q_{\alpha})
= 0 \label{j14}
\end{eqnarray}\ \\

\noindent Einstein equations:
\begin{eqnarray}
\fl \partial_0 \theta = -\frac{1}{3}\theta^2+2\omega^2-\frac{1}{2}(\mu+3 p)+j \label{ein00}\\
\fl 2/3 z_1 + \partial_2 \omega - \omega (r_2 - 2  \dot{u}_2) = 0\label{ein01}\\
\fl 2/3 z_2- \partial_1 \omega - \omega ( q_1 + 2  \dot{u}_1 ) = 0\label{ein02}\\
\fl 2/3 z_3- \omega n_{33} = 0\label{ein03}\\
\fl -\partial_1 r_2 +\partial_2 q_1 -r_1 r_2-q_1 q_2-2 r_2 q_1-2 r_3 n+n_{33} (r_3+q_3)-2 q_3 n =\nonumber \\
 -\partial_1 \dot{u}_2-\partial_2 \dot{u}_1-2 \dot{u}_1 \dot{u}_2+\dot{u}_1 q_2-\dot{u}_2 r_1\label{ein12}\\
\fl  -\partial_2 r_3 +\partial_3 q_2 +\partial_1 n -\partial_1 n_{33} -n (r_1-q_1)-2 q_1 n_{33}-r_2 r_3-q_2 q_3 -2 r_3 q_2= \nonumber \\
-\partial_2 \dot{u}_3-\partial_3 \dot{u}_2 +\dot{u}_1 (n_{33}-n) -2 \dot{u}_2 \dot{u}_3+\dot{u}_2 q_3-\dot{u}_3 r_2 \label{ein23}\\
\fl \partial_1 q_3 -\partial_3 r_1 +\partial_2 n_{33} -\partial_2 n -r_2 (2  n_{33}-n)-n q_2-r_1 r_3-q_1 q_3-2 r_1 q_3 =\nonumber  \\
-\partial_1 \dot{u}_3 -\partial_3 \dot{u}_1 -\dot{u}_1(2  \dot{u}_3+r_3)+\dot{u}_2(n- n_{33})+\dot{u}_3 q_1\label{ein13}\\
\fl -\partial_1 r_1 +\partial_1 q_1 +\partial_2 q_2 -\partial_3 r_3+q_2^2+r_3^2+\frac{1}{2} n_{33}^2-n n_{33}+r_1^2+q_1^2-r_2 q_2-r_3 q_3 = \nonumber \\
\frac{2}{9} \theta^2+\frac{8}{3} \omega^2-\frac{1}{3}(2 \mu- j)-\partial_1 \dot{u}_1-\dot{u}_1^2+\dot{u}_3 r_3-\dot{u}_2 q_2\label{ein11}\\
\fl -\partial_2 r_2+\partial_2 q_2-\partial_1 r_1+\partial_3 q_3 +q_3^2+r_1^2+\frac{1}{2} n_{33}^2-n n_{33}+r_2^2+q_2^2-r_1 q_1-r_3 q_3 =\nonumber  \\
\frac{2}{9} \theta^2+\frac{8}{3} \omega^2-\frac{1}{3}(2\mu-j)-\partial_2 \dot{u}_2-\dot{u}_2^2+\dot{u}_1 r_1-\dot{u}_3 q_3\label{ein22}\\
\fl -\partial_3 r_3 +\partial_3 q_3 +\partial_1 q_1 -\partial_2 r_2 +q_1^2+r_2^2-\frac{1}{2} n_{33}^2+r_3^2+q_3^2-r_1 q_1-r_2 q_2 = \nonumber \\
\frac{2}{9} \theta^2+\frac{1}{3}(2 \omega^2-2\mu+j)-\partial_3
\dot{u}_3 -\dot{u}_3^2+\dot{u}_2 r_2-\dot{u}_1 q_1\label{ein33}
\end{eqnarray}

\end{document}